\documentclass[fleqn,usenatbib,useAMS]{mnras}

\usepackage{newtxtext,newtxmath} 
\usepackage[T1]{fontenc}
\DeclareRobustCommand{\VAN}[3]{#2}
\let\VANthebibliography\thebibliography
\def\thebibliography{\DeclareRobustCommand{\VAN}[3]{##3}\VANthebibliography}

\usepackage{amsmath}	
\usepackage{graphicx}	
\usepackage{subcaption}
\usepackage{sirimacro}
\usepackage{xcolor}
\usepackage{multirow}
\usepackage{soul}
\def\simpropto{\lower.2ex\hbox{$\; \buildrel \propto \over \sim \;$}}
\def\ltsim{\lower.5ex\hbox{$\; \buildrel < \over \sim \;$}}
\def\gtsim{\lower.5ex\hbox{$\; \buildrel > \over \sim \;$}}

\definecolor{dark blue}{rgb}{0.00, 0.00, 0.55}
\definecolor{dark green}{rgb}{0.00, 0.39, 0.00}
\definecolor{dark red}{rgb}{0.55, 0.00, 0.00}

\newcommand{\eref}[1]{equation~(\ref{#1})}
\newcommand{\esref}[1]{equations~(\ref{#1})}
\newcommand{\fref}[1]{Fig.~\ref{#1}}
\newcommand{\fsref}[2]{Fig.~\ref{#1} and~\ref{#2}}
\newcommand{\fssref}[3]{Fig.~\ref{#1}, \ref{#2} and~\ref{#3}}

\newcommand{\sref}[1]{section~\ref{#1}}
\newcommand{\tref}[1]{Table~\ref{#1}}

\title{Extreme-value modelling of the brightest galaxies at $z\gtrsim9$} 

\author[C. Heather et al.]{
Cameron Heather,$^{1}$\thanks{E-mail:cameron.heather@warwick.ac.uk}
Teeraparb Chantavat,$^2$
Siri Chongchitnan,$^{1}$
and Joseph Silk$^{3,4,5}$
\\
$^{1}$Warwick Mathematics Institute, University of Warwick, Zeeman Building, Coventry CV4 7AL, UK\\
$^{2}$Institute for Fundamental Study, Naresuan University, Phitsanulok, 65000, Thailand\\
$^{3}$Institut d'Astrophysique de Paris, 98 bis Boulevard Arago, 75014, Paris, France \\
$^{4}$William H. Miller III Department of Physics and Astronomy,The Johns Hopkins University, Baltimore, MD 21218, USA\\
$^{5}$BIPAC, Department of Physics,University of Oxford, Keble Road, Oxford OX1 3RH, UK\\
}

\date{Accepted XXX. Received YYY; in original form ZZZ}

\pubyear{2024}

\begin{document}
\label{firstpage}

\pagerange{\pageref{firstpage}--\pageref{lastpage}}
\maketitle

\begin{abstract}
Data from the James Webb Space Telescope have revealed an intriguing population of bright galaxies at high redshifts. In this work, we use extreme-value statistics to calculate the distribution (in UV magnitude) of the brightest galaxies in the redshift range $9 \lesssim z \lesssim  16$. We combine the Generalised Extreme Value (GEV) approach with modelling of the galaxy luminosity function. We obtain predictions of the brightest galaxies for a suite of luminosity functions, including the Schechter and double power law functions, as well as a model parametrised by the stellar formation efficiency $f_*$. We find that the \textit{JWST} data is broadly consistent with $f_*$ of $5\%-10\%$, and that the brightest galaxy at $z\sim16$ will have  $M_{\rm UV}\approx -23.5^{0.8}_{0.4}$. If $f_*$ is dependent on halo mass, we predict $M_{\rm UV}\approx -22.5^{0.5}_{1.5}$ for such an object. We show that extreme-value statistics not only predicts the magnitude of the brightest galaxies at high redshifts, but may also be able to distinguish between  models of star formation in high-redshift galaxies.
\end{abstract}


\section{Introduction}

One of the primary objectives of contemporary astrophysics is to use the latest observational technologies to improve our understanding of the physics of galaxy formation and evolution \citep{Bromm_Yoshida2011, Stark2016, Ouchi_ea2020, Robertson2022}.  In pursuit of this goal, a new generation of multi-wavelength surveys has shed a new light on galaxies at unprecedentedly  high redshifts \citep{Gardner_ea2006, Laureijs_ea2011, ALMA2015, Spergel_ea2015, HERA2017, Marchetti_ea2017}.

One of the essential ingredients for studying galaxy evolution is the rest-frame ultraviolet (UV) luminosity function (LF), which characterises the density of galaxies within a given volume based on their luminosity.  The UV LF is well constrained within the redshift range of $2 \lesssim z \lesssim 10$ based on observations from the \textit{Hubble Space Telescope} (\textit{HST}) and \textit{Spitzer}, predating the launch of the \textit{James Webb Space Telescope} (\textit{JWST}) in 2021.  Many studies at that time suggest a rapid decline of the UV LF at the bright end and find that it is well described by the Schechter function \citep{Ellis_ea2013, Madau_Dickinson2014, Bouwen_ea2015, Bouwens_ea2021, Finkelstein_ea2015, Ishigaki_ea2018, Oesch_ea2018}.  However, some evidence suggests that the decline at the bright end of the UV LF is not as steep as predicted by the Schechter function.   \citep{Hathi_ea2012, Finkelstein_ea2013, Bowler_ea2014, Bowler_ea2020, McLeod_ea2024}.  Some authors have proposed that the bright end of the UV LF at $z \sim 4 - 7$ ($M_{\rm UV} < -24$) cannot be adequately explained by the Schechter function alone, suggesting the necessity for either a double power law or a modified Schechter function to better account for observational findings \citep{Ono_ea2018}.

The early release of \textit{JWST}/NIRCam data in July 2022 has enabled us to investigate the UV LF at redshifts beyond $10$, with the furthest candidate at approximately $z \sim 16$ through photometry \citep{Castellano_ea2022, Castellano_ea2023, Finkelstein_ea2022, Naidu_ea2022, Adams_ea2023, Atek_ea2023, Bradley_ea2023, Labbe_ea2023, Tacchella_ea2023, tacchella2023jades} , along with some spectroscopic redshift determinations \citep{Schaerer_ea2022, Arrabal-Haro_ea2023, Bunker_ea2023, Curti_ea2023, Curtis-Lake_ea2023, Heintz_ea2023a, Heintz_ea2023b, harikane2023pure}.  Instead of the rapid decline in UV LF described by the Schechter function at the bright end, recent data from \textit{JWST} indicates a more gradual decrease, better captured by a double power law \citep{Finkelstein_ea2022, Naidu_ea2022, Donnan_ea2023, donnan2024jwst}.  

The reasons for the deviation from a Schechter function to a double power law at high redshifts are varied and speculative. Explanations include: a high star formation efficiency at high redshifts \citep{harikane2023pure}, the top-heavy nature of the stellar initial mass function for Population III stars \citep{Finkelstein_ea2023, Bovill_ea2024, Lapi_ea2024}, the removal of interstellar dust through radiatively driven outflows during the initial stages of galaxy formation \citep{Ferrara_ea2023, Fiore_ea2023, Ziparo_ea2023},  short-term variation in dust attenuation  \citep{Mirocha_ea2023}, reduced negative or even positive feedback at early times  \citep{Silk_ea2024}, or a combination of these effects. 

High-redshift galaxies observed by the \textit{JWST} are expected to be amongst the most luminous within the field of view at their respective redshifts. Consequently, it is essential to account for any potential biases that may arise when interpreting the UV luminosity function from those luminous galaxies especially at the bright end of the UV LF. In other words, it is crucial to determine whether the UV-bright galaxies observed are representative of the extreme population from the tail end of the luminosity distribution.  In this study, we will address this question using \ii{extreme-value statistics} and derive a semi-analytical model of the luminosity distribution for the brightest galaxies at high redshifts and compare that with \ii{JWST} observation.  


A pioneering work on using extreme-value statistics to study bright galaxies was   \cite{bhavsar1985first}. They modelled the distribution of the brightest galaxies in groups and clusters at a redshift range $z < 3$ using what is now known as the Generalised Extreme-Value distribution (which we will discuss later in \sref{Sec:EVS}). Our work naturally follows their footsteps but extends the scope to a larger population of galaxies using a more sophisticated modelling of galaxy characteristics.

This paper will be organised as follows: in \sref{sec_LUM}, we will describe models of the galaxy luminosity function. Alongside the Schechter and the  double power law functions, we will introduce the star formation efficiency model, which is parametrised by the star formation efficiency $f_*$.  In \sref{Sec:EVS}, we will introduce a formalism for determining the distribution of the most luminous high-redshift galaxies using extreme-value statistics   and compare our theoretical model with the \ii{JWST} data. In \sref{sec:SFR}, we expand our model by introducing variations in the star formation efficiency with respect to halo mass. Our conclusions and discussion will be presented in \sref{sec:conclusionanddiscussion}.

We will assume a flat $\Lambda $CDM cosmology with $H_0 = 68.0\ \rm{km\  s^{-1}\ Mpc^{-1}} $, $\Omega_{\rm m} = 0.307 $, $\Omega_{\Lambda} = 0.693$, $\Omega_{\rm b} = 0.046$ and $\sigma_8 = 0.823$. 

\section{The Luminosity Function}
\label{sec_LUM}

The luminosity function gives the number density of galaxies per intrinsic luminosity interval $[L, L+\rm{d}L]$ at a given redshift. We will use the absolute UV magnitude, $M_{\rm UV}$, as a measure of luminosity, $L$, via the relation 
\begin{eqnarray}\label{eqn:Muv_L}
    M_{\rm UV} - M^*_{\rm UV} = -2.5\log_{10}\left(\frac{L}{L^*}\right),
\end{eqnarray}
where $M^*_{\rm UV}$ is a characteristic magnitude and $L^*$ is a characteristic luminosity which are introduced to allow the functions to be fitted to some observed data.

We now discuss two approaches to obtain the luminosity function of high-redshift galaxies. The first method is to fit the luminosity function to observational data using the double power law and Schechter functions. The second method uses a halo mass function combined with a model of the stellar formation rate.

\subsection{Double Power Law and Schechter Models}\label{subsec:obs}

The double power law function is used to model the luminosity of galaxy groups and clusters (\cite{holmberg1969study}, \cite{tempel2009anatomy}) and of radio galaxies and quasars (\cite{dunlop1990redshift}).
We take the form of the double power law luminosity function from ~\cite{harikane2023pure}: 
\begin{multline}\label{eqn:dpl}
    \frac{\D n}{\D M_{\rm UV}} = \frac{\ln10}{2.5}\varphi^{*}  \\
     \times\left[10^{0.4(\tilde{\alpha} + 1)(M_{\rm UV} - M^*_{\rm UV})}+10^{0.4(\tilde{\beta} + 1)(M_{\rm UV} - M^*_{\rm UV})}\right]^{- 1},
\end{multline}
where the parameters are given by \citep{harikane2023comprehensive}:
\begin{eqnarray*}
    M^*_{\rm UV} &=& -0.09(z - 9) - 19.33, \\
    \log\varphi^* &=& -0.28(z - 9) - 3.50, \\
    \tilde{\alpha} &=& -2.10, \\
    \tilde{\beta} &=& 0.15(z - 9) - 3.27.
\end{eqnarray*}
\fref{fig:harikane} shows this luminosity function  (black solid lines, labelled DPL) plotted against magnitude $M_{\rm UV}$, along with the other luminosity functions that we will be exploring in this work. 

Next, the Schechter model \citep{schechter1976analytic} for the luminosity function is given by
\begin{eqnarray} \label{eqn:sch_lum}
    \frac{\D n}{\D L} = \varphi^*\left(\frac{L}{L^*}\right)^{\tilde{\alpha}} \exp\Big(-L/L^*\Big),
\end{eqnarray}
where $\varphi^*$ and $\tilde{\alpha}$ are the model parameters.  
Combining this with \eref{eqn:Muv_L}, the Schechter luminosity function can be expressed as
\begin{multline} \label{eqn:sch_mag}
    \frac{\D n}{\D M_{\rm UV}} = \frac{\ln10}{2.5}\varphi^{*} \left(-10^{0.4(M_{\rm UV} - M^*_{\rm UV})(\tilde{\alpha} + 1)}\right)\\
    \times \exp\left[-10^{0.4(M_{\rm UV} - M^*_{\rm UV})}\right],
\end{multline}
where the parameters are given by \citep{harikane2023comprehensive}:
\begin{eqnarray*}
    M^{*}_{\rm UV} &=& -0.32(z-9) - 21.24, \\
    \log(\varphi^*) &=& -0.08(z-9) - 4.83, \\
    \tilde{\alpha} &=& -2.35.
\end{eqnarray*}

We also plot the Schechter function against magnitude in \fref{fig:harikane} (black dashed lines). 



\subsection{The Star Formation Efficiency Model}\label{subsec:fstar}

In this model, we start by  modelling the abundance of bright galaxies using a halo mass function coupled with some model of the stellar formation rate (SFR). A particularly simple model of the SFR (previously used in  \cite{harikane2023pure}) is given by:
\begin{eqnarray}\label{eqn:SFR}
    {\rm SFR} = f_{*} \times f_{\rm b} \times \frac{\D M_{\rm h}}{\D t}(M_{\rm h}, z),
\end{eqnarray}
where $f_{*}$ is the \emph{star formation efficiency}, $f_{\rm b} \equiv \Omega_{\rm b}/ \Omega_{\rm m}$ is the cosmic baryon fraction, and  $\D M_{\rm h}/\D t$ is the matter accretion rate for a halo of mass $M_{\rm h}$. We will use the model of the accretion rate given in \cite{behroozi2015simple}.

In this section, we will take $f_*$ to be constant, taking the values $0.02, 0.05, 0.15, 0.4, 1$ (\sref{sec:SFR} discusses a more sophisticated model of $f_*$). We then  convert the $\rm SFR$ into UV luminosity, $L_{\rm UV}$, via 
\begin{eqnarray}\label{eqn:LUV}
    L_{\rm UV}({\rm erg\ s}^{-1} {\rm Hz}^{-1}) = {\rm SFR}(M_{\odot}{\rm yr}^{-1})/(1.15 \times 10^{-28}).
\end{eqnarray} Combining \esref{eqn:SFR} and~(\ref{eqn:LUV}), we obtain an equation for luminosity in terms of halo mass, $M_{\rm h}$.

Next, luminosity is converted to UV magnitude using values taken from \cite{oke1983secondary}, such that \eref{eqn:Muv_L} becomes
\begin{eqnarray}\label{oke}
    M_{\rm UV} = -2.5 \log_{10}\left(\frac{L_{\rm UV}}{{\rm erg\ s}^
    {-1} {\rm Hz}^{-1}}\right) + 51.6.
\end{eqnarray}
This gives us the following relation between the UV magnitude and halo mass.
\begin{multline}\label{eqn:Muv}
    M_{\rm UV}(M_{\rm h}, z) = -2.5 \log_{10}\left(\dot{M}_{\rm h}\right) -2.5 \log_{10}\left(\frac{ f_* \times f_{\rm b}}{1.15 \times 10^{-28}}\right) + 51.6
\end{multline}
Taking the derivative with respect to $M_{\rm h}$, we obtain
\begin{eqnarray}\label{dMdM}
     \frac{\D M_{\rm UV}}{\D M_{\rm h}} = -\frac{2.5}{\ln(10)}\frac{\ddot{M}_{\rm h}}{\dot{M}_{\rm h}^2}.
\end{eqnarray}

We now introduce the halo mass function, $\D n/\D M_{\rm h}$, defined as the number of dark matter haloes per unit mass interval $[M_{\rm h}, M_{\rm h}+\D M_{\rm h}]$. 
\begin{eqnarray} \label{eqn:tinker}
    \frac{\D n}{\D M_{\rm h}} = f(\sigma)\frac{\bar{\rho}_{\rm m}}{M_{\rm h}}\frac{\D \ln \sigma^{-1}}{\D M_{\rm h}}.
\end{eqnarray}
Following \cite{behroozi2015simple}, we will use the Tinker mass function \citep{tinker2008toward} in which 
\begin{eqnarray}
    f(\sigma) = A\left[{\left(\frac{\sigma}{b}\right)}^{-a}+1\right]e^{-c/\sigma^2},
\end{eqnarray}
where $\sigma$ is the variance of the smoothed linear density field, and the parameters $A$, $b$ and $c$ depend on redshift.
The luminosity function can then be expressed via the chain rule as:
\begin{align} 
    \hskip 2 cm\frac{\D n}{\D M_{\rm UV}} &= \frac{\D n}{\D M_{\rm h}}\left|\frac{\D M_{\rm h}}{\D M_{\rm UV}}\right|\label{eqn:lum_func}\\
    &=\frac{\D n}{\D M_{\rm h}} \times 0.4\ln(10)\left|\frac{\dot{M}_{\rm h}^2}{\ddot{M}_{\rm h}} \right|.\label{eqn:newlum}
\end{align}


In \eref{eqn:newlum}, there appears to be no direct dependence on $f_*$. However, the $f_*$ dependence does arise through the conversion formulae in \esref{eqn:LUV} and~(\ref{eqn:Muv}). 



\begin{figure*}
    \centering
    \includegraphics[width =  \textwidth]{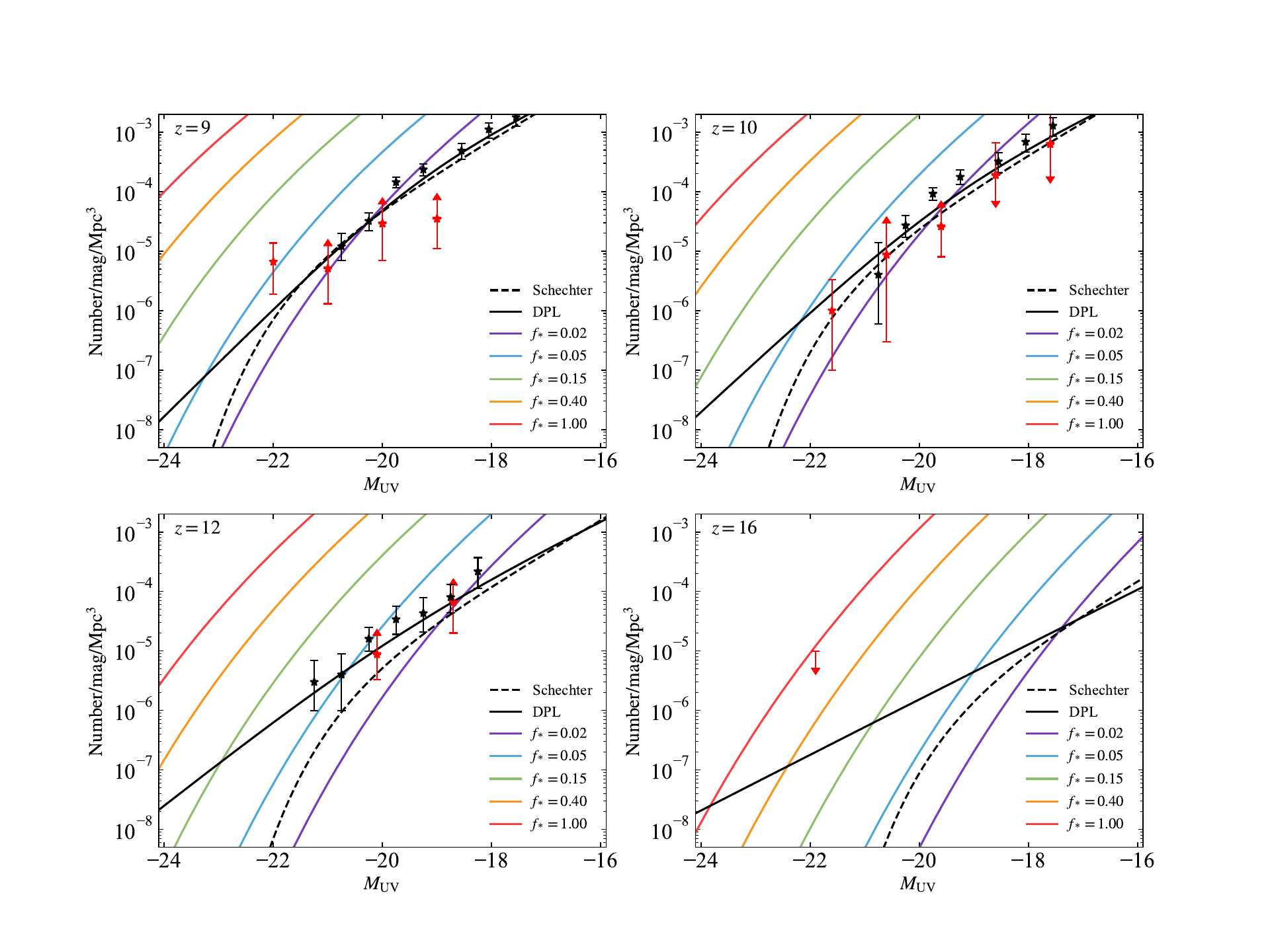}
    \caption{ Comparison of the luminosity functions $\D n/\D M_{\rm UV}$ at $z = 9$ (upper-left), $z = 10$ (upper-right), $z = 12$ (lower-left) and $z = 16$ (lower-right). The black solid and dashed curves are the double power law and Schechter luminosity functions respectively, calculated in \sref{subsec:obs}. The colourful curves correspond to the $f_*$ model of the luminosity function in \sref{subsec:fstar} (see \eref{eqn:newlum}). The colourful curves correspond to  $f_* = 0.02, 0.05, 0.15, 0.4, 1$, with amplitudes matching those of   \protect\cite{harikane2023pure}. The red data points correspond to spectroscopic data from \protect\cite{harikane2023pure}, and the black points correspond to photometric data from \protect\cite{donnan2024jwst}.} \label{fig:harikane}
\end{figure*}

We plot the various luminosity functions calculated in this section at four redshifts $(z = 9, 10, 12, 16)$ in \fref{fig:harikane}, where $f_*$ is varied between 0.02 and 1 (the curves for the $f_*$ model are normalised to match those of \cite{harikane2023pure}). We see that the Schechter luminosity function is comparable to the $f_*$ model with $f_*$ roughly between 0.02 and 0.05. On the other hand, the double power law function is flatter and  predicts a higher number of brighter galaxies than the other models.

\section{Extreme-Value Modelling of the Brightest Galaxies}\label{Sec:EVS}

Our goal is to obtain a statistical description of the most extreme luminosities of high-redshift galaxies. We will do this using extreme-value statistics, which has previously been used in astrophysics to find the brightest galaxies in clusters by~\cite{bhavsar1985first}, quantify the abundances of the most massive Pop III stars (\cite{Chantavat_ea2023}), predict the most massive galaxy clusters \citep{davis2011most, waizmann2012application, chongchitnan2012primordial} and estimate the abundances of extreme primordial black holes and extreme-spin primordial black holes  \citep{Chongchitnan_ea2021, chongchitnanspin}. 

In particular, we will be working with the generalised extreme-value (GEV) approach, also known as the block maxima method (\cite{gumbel1958statistics}).
In this method, we divide the galaxy population  in a given redshift bin into $N$ distinct blocks, from which we identify the brightest galaxy in each block using a probability density function outlined below. Analogous to the Central Limit Theorem, these brightest galaxies will have a large-$N$ limit distribution corresponding to one of the GEV distributions (\eref{eqn:GEV}). 

We summarise the GEV pipeline here (more details can be found in \cite{white1979hierarchy, davis2011most}). First, we calculate the number density $n(<M_{\rm UV})$ of galaxies with UV  magnitude less than $M_{\rm UV}$ as
\begin{eqnarray}\label{eqn:num_dens}
    n(<M_{\rm UV},z) = \int_{-\infty}^{M_{\rm UV}} \frac{\D n}{\D M_{\rm UV}'} \D M_{\rm UV}'.
\end{eqnarray}
Note that, by the definition of $M_{\rm UV}$, \eref{eqn:num_dens} counts the \textit{brightest} galaxies. The redshift dependence can then be integrated out to obtain the number count of the brightest galaxies in the redshift interval $[z_0,z_1]$:
\begin{eqnarray}\label{eqn:num_count}
    n(<M_{\rm UV}) = \int_{z_0}^{z_1} n(<M_{\rm UV},z) \D z.
\end{eqnarray}

Next, to predict the brightest magnitude that galaxies could have at high redshifts, we consider the probability $P_0$ that no galaxies in a given volume  exceeds the extreme brightness $M_{\rm UV}$. In our calculation for the volume $V$, we take $f_{\rm sky} = 1$ in order to obtain  generic predictions on the extreme luminosities.  The probability distribution $P_0(M_{\rm UV})$ can be modelled as a Poisson distribution with the following cumulative distribution function (cdf):
\begin{eqnarray}\label{eqn:Poisson}
    P_0(M_{\rm UV}) = \exp(-n(<M_{\rm UV})V),
\end{eqnarray}
Differentiating this with respect to $M_{\rm UV}$, we obtain the probability density function (pdf):
\begin{eqnarray}\label{eqn:Pois_pdf}
   \frac{\D P_0}{\D M_{\rm UV}} = -V\frac{\D n(<M_{\rm UV})}{\D M_{\rm UV}}\exp(-n(<M_{\rm UV})V),
\end{eqnarray}

 The  Fisher-Tippett-Gnedenko theorem implies  that in the large-$N$ limit, the cdf (\eref{eqn:Poisson}) approaches the GEV distribution given by:
\begin{eqnarray}\label{eqn:GEV}
    G(M_{\rm UV}) = 
    \begin{cases}
        \exp\left[ -(1+\gamma y)^{-1/\gamma}\right] & (\gamma \neq 0), \\
        \exp\left[ -e^{-y}\right] & (\gamma = 0),
    \end{cases}
\end{eqnarray}
where $ y := (M_{\rm UV} - \alpha)/ \beta$, with $\alpha$ describing the location of the peak, and $\beta$ describing the scale of the pdf. The sign of the parameter $\gamma$ determines the GEV type, with $\gamma = 0$, $\gamma>0$ and $\gamma<0,$ corresponding to the Gumbel, Fr\'echet and Weibull distributions respectively.

We can express the GEV parameter $\alpha, \beta, \gamma$ in terms of astrophysical parameters by Taylor-expanding the Poisson distribution $P_0(M_{\rm UV})$ and the GEV distribution $G(M_{\rm UV})$ around the peak of the pdf at $M_{\rm peak}$ to cubic order. By equating the coefficients, we find 

\begin{eqnarray}\label{eqn:params}
\begin{aligned}
    \gamma &= n(<M_{\rm peak})V - 1, \\\\
    \alpha & = M_{\rm peak} - \frac{\beta}{\gamma}\left({(1+\gamma)}^{-\gamma}-1\right), 
\end{aligned} 
\begin{aligned}
    \beta &= \frac{{(1+\gamma)}^{1+\gamma}}{\displaystyle\frac{\D n}{\D M_{\rm UV}}\Big|_{M_{\rm peak}}V}.
\end{aligned}
\end{eqnarray}

While these equations provide a good approximation to the values $\gamma$, $\alpha$ and $\beta$, using a numerical approximation produces a closer fit to the Poisson pdf (\eref{eqn:Pois_pdf}), which we use to plot the GEV distribution in \fref{fig:pdf}. This method assumes the pdfs are well described by the Gumbel distribution, taking $\gamma \approx 0$.


\subsection{Extreme-value pdf of the brightest galaxies}

\begin{figure*} 
    \centering
    \includegraphics[width = \textwidth ]{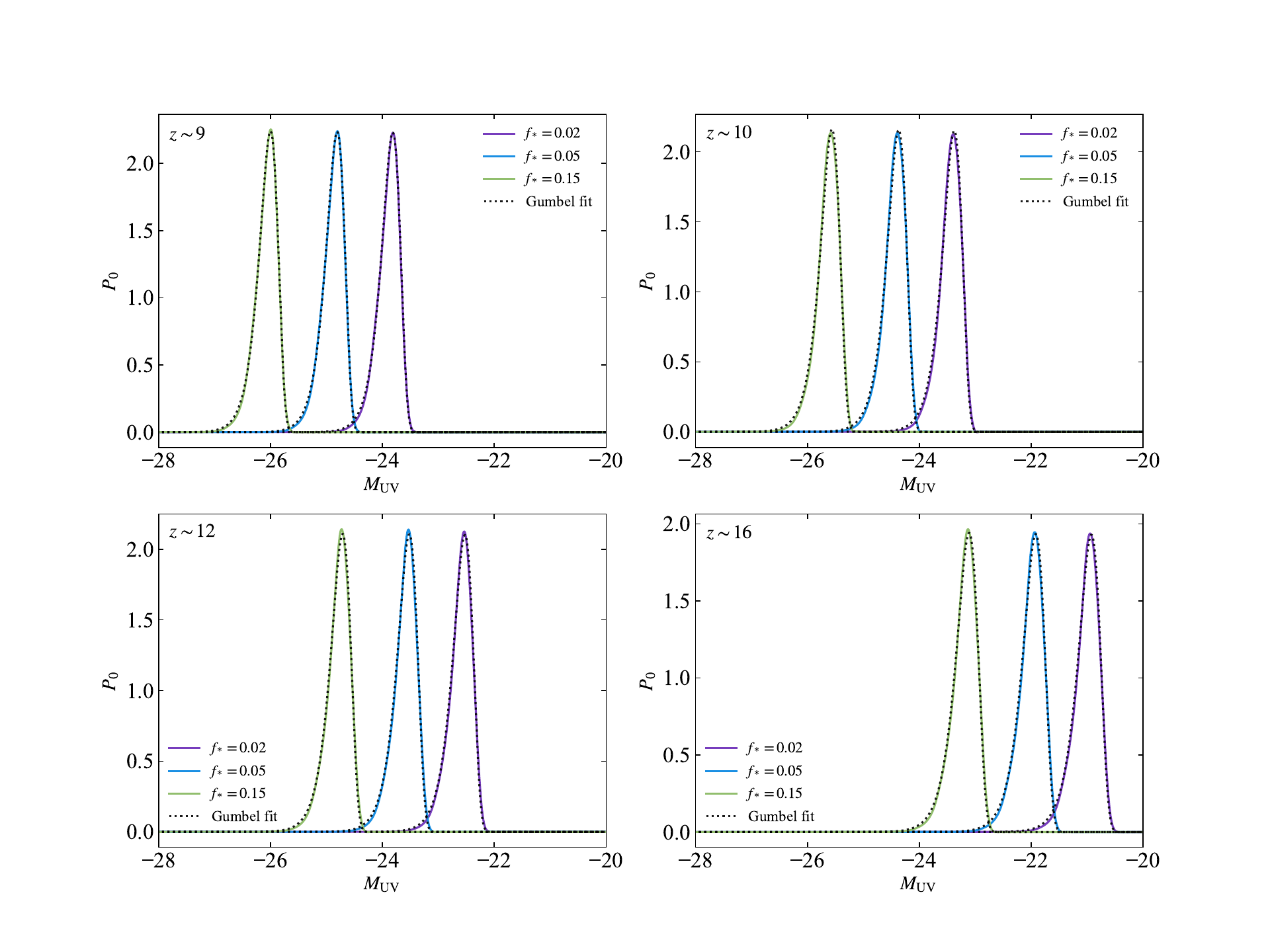}
    \caption{The probability density functions for the most luminous galaxies calculated in \sref{subsec:fstar} for 4 redshift bins: $z=[9, 10]$ (upper-left), $[10, 11]$ (upper-right), $[12, 13]$ (lower-left) and $[16, 17]$ (lower-right). The solid coloured lines are the pdfs derived from the Poisson distribution (\eref{eqn:Pois_pdf}), assuming 
    3 values for the star formation efficiency: $f_* = 0.02$ (purple), $0.05$ (blue) and $0.15$ (green). The dashed lines are the Gumbel distribution (\eref{eqn:GEV}) with parameters $\alpha, \beta$ given in \tref{tab:full}.}\label{fig:pdf}
\end{figure*}

We plot the pdf for the distribution of the brightest galaxies in \fref{fig:pdf} for 4 redshift bins: $z\in[9, 10]$, $[10,11]
$, $[12, 13]$ and $[16,17]$. The solid lines are pdfs calculated using \eref{eqn:Pois_pdf}, whilst the dashed lines are the numerical fits assuming that the pdfs are Gumbel type ($\gamma=0$ in \eref{eqn:GEV}). 
The parameters $\alpha$ and $\beta$ for the Gumbel fit are given in \tref{tab:full}. Note that in each bin, $\alpha$ is very close to the magnitude of $M_{\rm UV}$ where the peak of the pdf occurs.

From \fref{fig:pdf}, we see that at higher redshifts, the pdfs shift to higher values of $M_{\rm UV}$ (meaning that the extreme galaxies are less bright). We also observe that increasing $f_*$ produces brighter extreme galaxies. In particular, increasing $f_*$ from $0.02$ to $0.15$ results in a pdf peak shift of $\Delta M_{\rm UV}\approx-2$.


\subsection{Comparison with \textit{JWST} data}\label{sec:bright}

\begin{figure*} 
    \centering
    \includegraphics[width = 0.49\textwidth ]{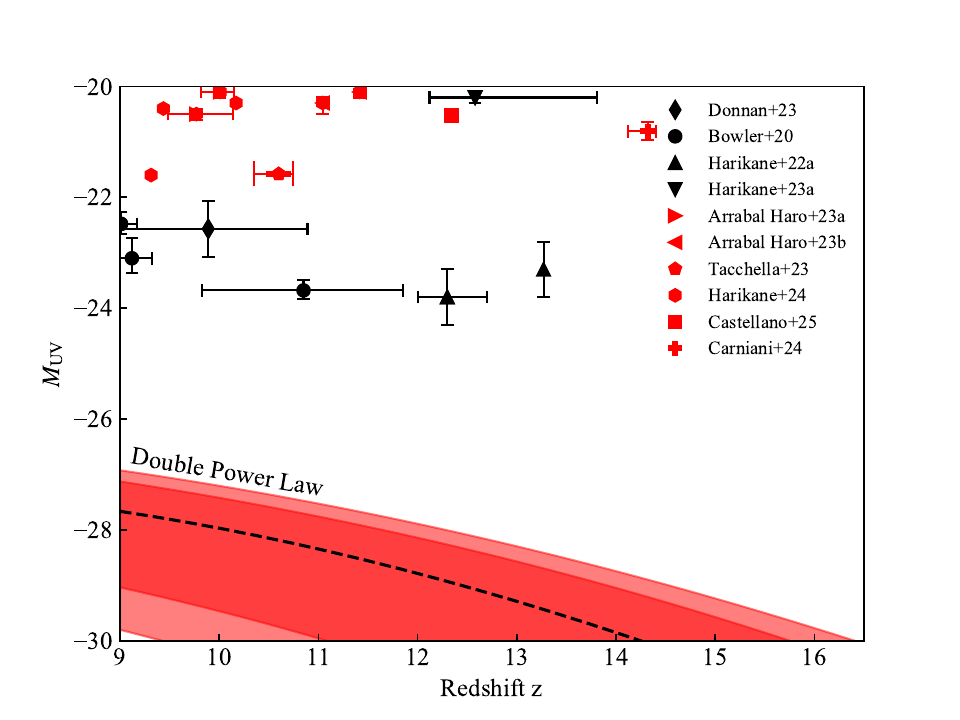}
    \includegraphics[width = 0.49\textwidth ]{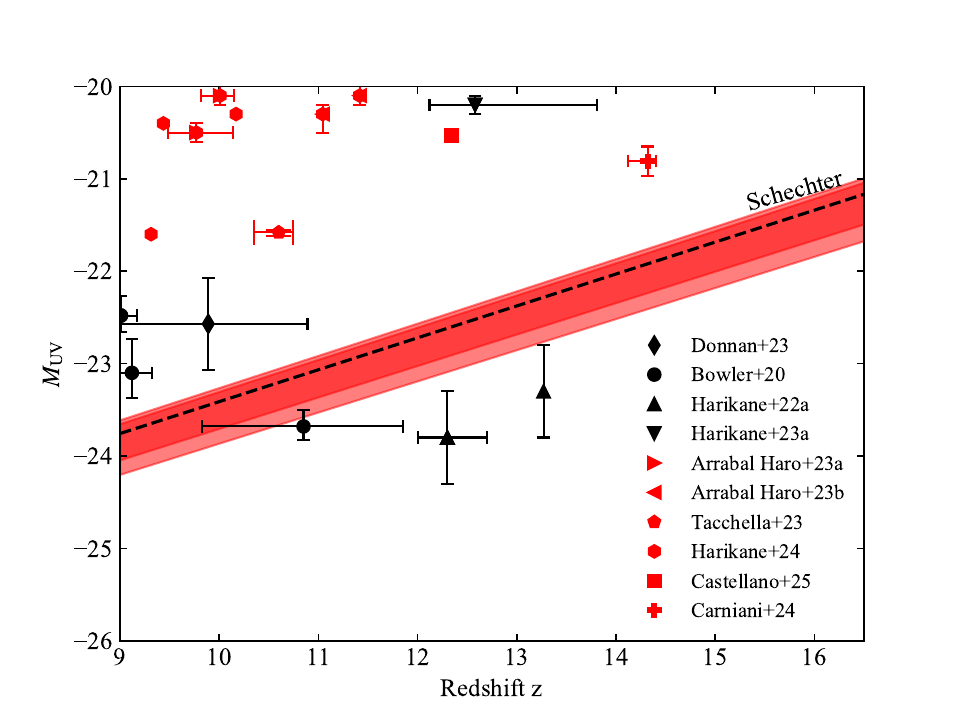}
    \caption{The peak magnitudes $M_{\rm peak}$ (dashed lines) of the extreme-value (Gumbel) pdf for $M_{\rm UV}$ plotted as a function of redshift, assuming the double power law luminosity function from \eref{eqn:dpl} (left) and the Schechter luminosity function from \eref{eqn:sch_mag} (right).  The darker/lighter regions correspond to the $95\super{th}$ and $99\super{th}$ percentiles of the pdf. The data points are based off the most extreme samples from previous studies (\protect\cite{Donnan_ea2023}, \protect\cite{Bowler_ea2020}, \protect\cite{harikane2022search}, \protect\cite{harikane2023comprehensive}, \protect\cite{Arrabal-Haro_ea2023}, \protect\cite{haro2023spectroscopic}, \protect\cite{tacchella2023jades}, \protect\cite{harikane2023pure}, \protect\cite{castellano24}, \protect\cite{carniani24}).}  \label{fig:evs}
\end{figure*}

\begin{figure} 
    \centering
    \includegraphics[width = 0.48\textwidth ]{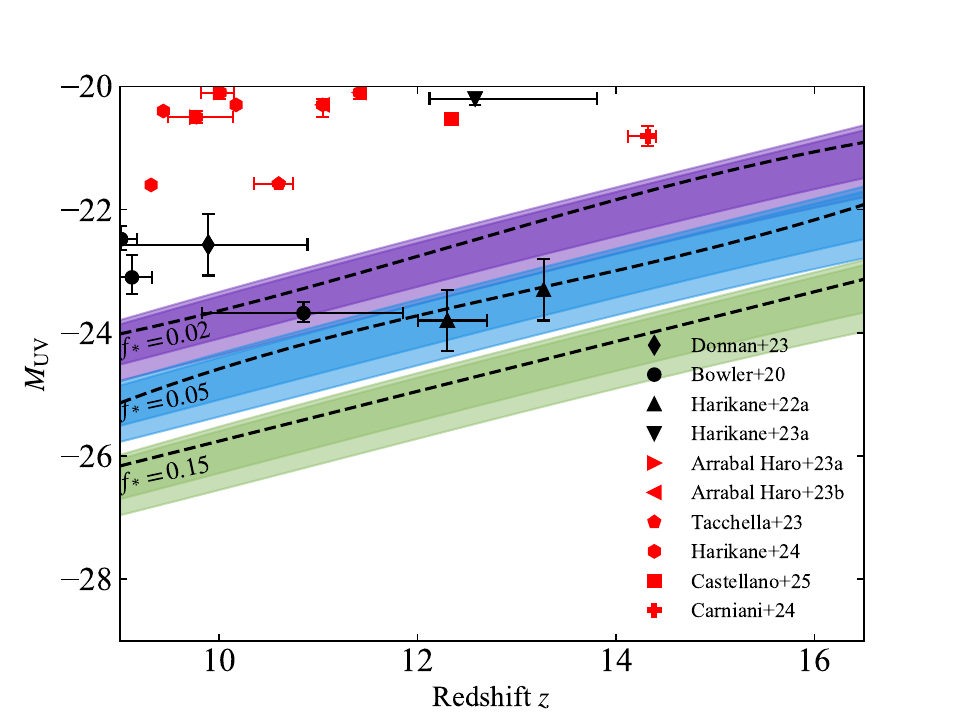}
    \caption{Extreme-value pdf profile (same as \fref{fig:evs}) plotted as a function of redshift, for the $f_*$ model  of the luminosity function in \sref{subsec:fstar} (\eref{eqn:newlum}). The values for the star formation efficiency shown are  $f_* = 0.02$ (purple), $0.05$ (blue) and $0.15$ (green).}\label{fig:evs_f}
\end{figure}

We now study the redshift dependence of the extreme-value predictions more closely. In \fsref{fig:evs}{fig:evs_f}, we plot the peak of the Gumbel pdf for the brightest galaxies over the redshift range $ z = 9-16$ (dashed lines). We also indicate the $95\super{th}$ and $99\super{th}$ percentiles of the pdfs (shaded dark and light colours). \fref{fig:evs} shows the extreme-value predictions for $M_{\rm UV}$ for the double power law and Schechter models. \fref{fig:evs_f} shows the same predictions for the $f_*$ model with $f_*=0.02, 0.05$ and $0.15$.

On these plots, we have included data from the \textit{JWST} Early Release Science programs CEERS and GLASS. In particular, we have selected the brightest galaxies from a collection of recent studies on these galaxies (\protect\cite{Donnan_ea2023}, \protect\cite{Bowler_ea2020}, \protect\cite{harikane2022search}, \protect\cite{harikane2023comprehensive}, \protect\cite{Arrabal-Haro_ea2023}, \protect\cite{haro2023spectroscopic}, \protect\cite{tacchella2023jades}, \protect\cite{harikane2023pure}).  Most of these studies obtained the photometric redshifts for the galaxies using the EAZY or PROSPECTOR pipeline. While the redshifts have not been spectroscopically confirmed, the papers have published redshift uncertainties, which we have displayed as error bars. We also include spectroscopically confirmed galaxies from \protect\cite{Arrabal-Haro_ea2023}, \protect\cite{haro2023spectroscopic}, \protect\cite{tacchella2023jades}, \protect\cite{harikane2023pure}, \protect\cite{castellano24}, \protect\cite{carniani24}.



\subsubsection{Double power law and Schechter models}

In \fref{fig:evs}, we note that neither the double power law or Schechter model produces a satisfactory prediction for the peak magnitudes of the brightest galaxies. We discuss these separately below.

The double power law model (left panel) overestimates the brightness of extreme galaxies. For example, for $z>14.5$, the model predicts extreme brightness of $M_{\rm UV}\lesssim -30$. More so, it gives an unphysical prediction that the most luminous galaxies are \ii{brighter} at higher redshifts. We note that the unphysical trend can be corrected by an ad-hoc change in the sign $d\tilde{\beta}/dz$ in \eref{eqn:dpl}. We will revisit this point in the discussion section. 
 
Next, using the Schechter model (right panel), we find that the width of the pdf is much smaller. As a result, the coloured band barely contains any of the the extreme data points, and, more worryingly, we also see observed galaxies that are brighter than the predicted extreme brightness.


\subsubsection{The $f_*$ model}

In \fref{fig:evs_f} three of the star formation efficiency models have been plotted, $f_* = 0.02, 0.05, 0.15$. The model with $f_*=0.02$ is comparable to the Schechter model discussed above. The model with $f_* = 0.05$ (blue band) appears to be consistent with the data points, with the implication that these galaxies are expected to be the brightest we would observe. The model with $f_* = 0.15$ is also consistent with the data, in the sense that all the data points correspond to galaxies that are less bright than the extreme-value prediction. 

The extreme-value modelling allows us to extrapolate to higher redshifts. For example, assuming $f_* = 0.15$, at $z\sim 16$ we expect to see galaxies no brighter than $M_{\rm UV}\sim -23.5^{0.8}_{0.4}$.  Future observations can be added to such a plot, giving us additional constraints  on the star formation efficiency $f_*$. 

\section{Extending the  $\lowercase{f}_*$ model}\label{sec:SFR}

A natural extension to the $f_*$ model is to consider the case when the star formation efficiency is not a constant. Our goal is to obtain the extreme-value pdf for the brightest galaxies in this model.

\begin{figure}
\centering
\includegraphics[width =0.48\textwidth]{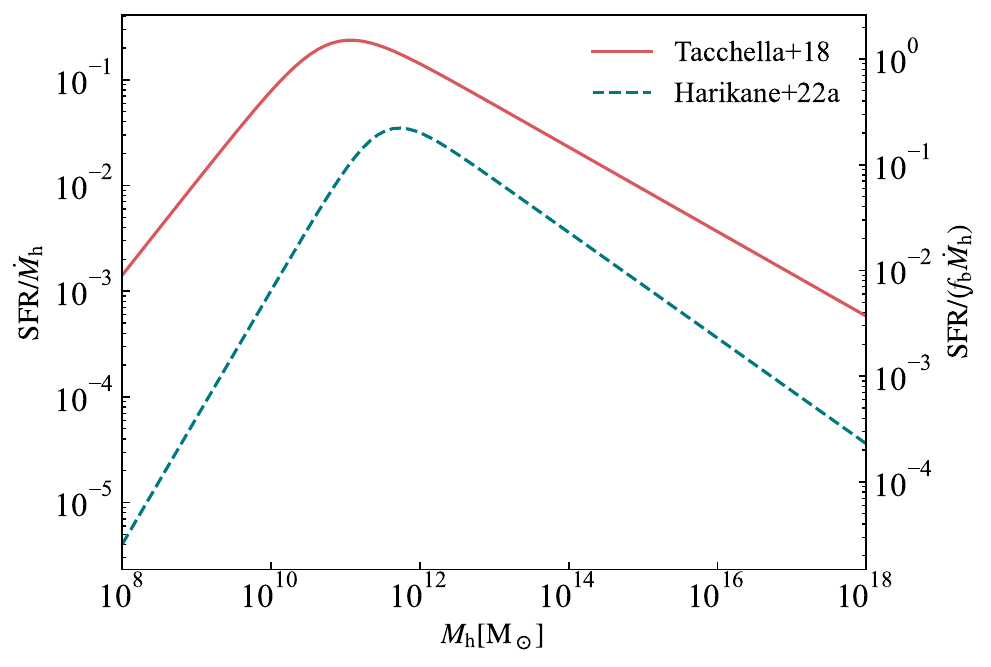}
\caption{Comparison of two stellar efficiency functions $f_*$ as a function of halo mass $M_{\rm h}$. The double-power law form is given in  \eref{eqn:fstar_mass}, with parameter values in \esref{Tmodel} and~(\ref{Hmodel}) for the Tacchella (pink) and Harikane (teal) models respectively.}
\label{fig:SFR}
\end{figure}

One such model is that of \cite{tacchella2018redshift} in which $f_*$ has a dependence on the halo mass $M_{\rm h}$ in the form of the double power law function: 
\begin{eqnarray}\label{eqn:fstar_mass}
    f_* = \frac{\rm SFR}{\dot{M}_{\rm h}} = 2\epsilon_0\left[\left(\frac{M_{\rm h}}{M_{\rm c}}\right)^{-\mu}+\left(\frac{M_{\rm h}}{M_{\rm c}}\right)^{\nu}\right]^{-1},
\end{eqnarray}
where $\epsilon_0$ is a normalisation constant, $M_{\rm c}$ is a characteristic mass where the efficiency is equal to $\epsilon_0$, and $\mu$ and $\nu$ are slopes that determine the behaviour at high and low masses. This equation has been calibrated at $z =4$, with parameter values: 
\ba\text{Tacchella model: }(\epsilon_0, M_{\rm c}, \mu, \nu)=(0.22, 6.3\times10^{10}\;{\rm M _{\odot}}, 0.89, 0.4).\lab{Tmodel}\ea
We also see the same form in \cite{harikane2022goldrush} with  parameter values:
\ba\text{Harikane model: }(\epsilon_0, M_{\rm c}, \mu, \nu)=(3.2\times10^{-2},10^{11.5}\;{\rm M _{\odot}}, 1.2, 0.5).\lab{Hmodel}\ea
These two models of $f_*$ are plotted in \fref{fig:SFR}.

We can follow the same method as \sref{subsec:fstar} to find the luminosity function for this mass-dependent $f_*$, by substituting \eref{eqn:fstar_mass} into the working. In particular, our equation for $M_{\rm UV}(M_{\rm h},z)$ (\eref{eqn:Muv}) is now
\begin{multline}\label{eqn:Muv_ext}
    M_{\rm UV}(M_{\rm h}, z) = -2.5 \log_{10}\left(\dot{M}_{\rm h}\right) +2.5\log_{10}\left(M_*^{-\mu}+M_*^{\nu}\right)\\ -2.5 \log_{10}\left(\frac{ 2\epsilon_0}{K_{\rm UV}}\right) + 51.6,
\end{multline}
where $M_* = M_{\rm h}/M_{\rm c}$. We can then take the derivative with respect to $M_{\rm h}$ and obtain the luminosity function:
\begin{eqnarray}\label{eqn:dndMfstar}
    \frac{\D n}{\D M_{\rm UV}} = \frac{\D n}{\D M_{\rm h}} \times 0.4\ln(10)\left|\frac{\ddot{M}_{\rm h}}{\dot{M}_{\rm h}^2} + \frac{\mu{M_{*}}^{-\mu} - \nu{M_{*}}^{\nu}}{M_{\rm h}\left({M_{*}}^{-\mu} +{M_{*}}^{\nu}\right)}\right|^{-1}.
\end{eqnarray}

\begin{figure}
    \centering
    \includegraphics[width =0.48\textwidth]{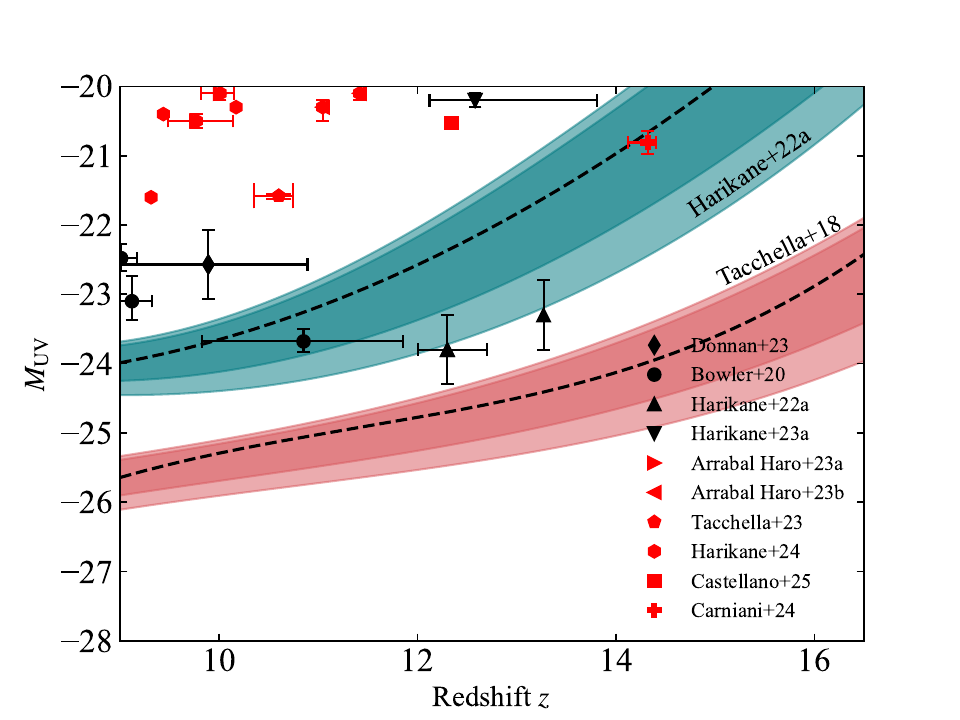}
    \caption{Extreme-value pdf profile (same as \fref{fig:evs}) plotted as a function of redshift, for the halo mass-dependent $f_*$ model (\eref{eqn:fstar_mass}). The teal (upper) band corresponds to the  Harikane model (\eref{Hmodel})  whilst the pink (lower) band is the Tacchella model (\eref{Tmodel}).}\label{fig:EVS_ext}
\end{figure}

From this, we introduce the extreme-value methodology from \sref{Sec:EVS}, where we find the peak magnitude for each redshift bin, which we then model using a GEV distribution. We plot the peak of Gumbel pdf, with the $95\super{th}$ and $99\super{th}$ percentile bands, varied over our redshift range, in \fref{fig:EVS_ext}.

The main difference between \fsref{fig:evs_f}{fig:EVS_ext} is the upward curvature of the bands at high redshifts. We also see that the Tacchella model gives a similar extreme-value prediction to the $f_* = 0.15$ model. At $z\sim16$, the Tacchella curve predicts an extreme brightness of $M_{\rm UV} = -22.5^{+0.5}_{-1.5}$, which is less bright than what the $f_* = 0.15$ model predicts. The confidence interval in the Tacchella model is also wider at the high-$z$ end of the graph.

The Harikane model, on the other hand, predicts a more conservative set of extreme brightnesses, with some observed bright galaxies barely grazing the $99\super{th}$ percentile band. Around  $z\sim10$, the extreme-value predictions in the Harikane model is comparable to that with constant $f_*$ model with $f_*$ of a few percent. However, the band curves up rapidly towards higher redshifts. Finally, like the Tacchella model, the band is also much wider than the constant $f_*$ model.

We conclude that both the Tacchella and the Harikane models are consistent with \textit{JWST} observations. However, considering the conservative volume coverage of \textit{JWST}, the Harikane model will require modification to remain consistent with the observations of the brightest  galaxies at $z\gtrsim9$.



\section{Conclusion and discussion}
\label{sec:conclusionanddiscussion}

We have applied extreme-value statistics to predict how bright the most luminous galaxies at high redshifts could be. In particular, we derived the probability distribution (in the UV magnitude $M_{\rm UV}$) of the brightest galaxies at redshifts $z\sim 9-16$. We found that such a distribution is well approximated by the Gumbel distribution. We studied a number of models of the galaxy luminosity function and derived the corresponding extreme-value distributions as a function of redshift.  Our main results are shown in \fsref{fig:evs_f}{fig:EVS_ext}, in which our theoretical results are compared with data from the \textit{JWST}.

Further summary and discussion points are given below. 

\begin{itemize}
\item \ii{The Generalised Extreme-Value (GEV) formalism} (\eref{eqn:GEV}) was used to calculate pdf of the brightest galaxies. We find that the GEV parameter $\gamma$ (see \tref{tab:full}) is so small that the extreme-value pdf for $M_{\rm UV}$ is well approximated by the Gumbel distribution, as shown in \fref{fig:pdf}. The pdf profile (along with the $95\super{th}$ and $99\super{th}$ percentiles) are shown in \fssref{fig:evs}{fig:evs_f}{fig:EVS_ext}, assuming different galaxy luminosity functions as shown in \fref{fig:harikane}.

\mmm

    \item \ii{The Schechter model} of the luminosity function  has previously been used to model the luminosity of galaxies at $z \sim 5$. The parameters used in this work (\eref{eqn:sch_lum}) were calibrated using observations at $z \sim 9$. We found that the resulting extreme-value pdf (right panel in \fref{fig:evs}) appears to be in tension with the brightest \textit{JWST} galaxies at $z\gtrsim10$. 
    

\mmm

    \item \ii{The double power law model} (\eref{eqn:dpl}), on the other hand, resulted in an unphysical extreme-value prediction (left panel in \fref{fig:evs}), where extreme galaxies appear brighter at higher redshifts. This might be due to the limited accuracy of the parametrisation (\eref{eqn:dpl}). Indeed, we found an ad-hoc correction by switching the sign of $\D\tilde{\beta}/\D z$ to negative.
    
    \cite{donnan2024jwst} gave new constraints on the redshift evolution of the double power law luminosity function. We use their data in place of the parametrisation (\eref{eqn:dpl}), and obtained the extreme-value pdf profile shown in \fref{fig:evs_dpl_new}, which appears to be consistent with data. We note that in their work, $\D\tilde{\beta}/\D z$ is indeed negative. Since the parameter $\tilde{\beta}$ describes the bright-end slope of the curve, we conclude that a physical double power law parametrisation must satisfy the condition $\D\tilde{\beta}/\D z<0.$


\begin{figure} 
    \centering
    \includegraphics[width = 0.48\textwidth ]{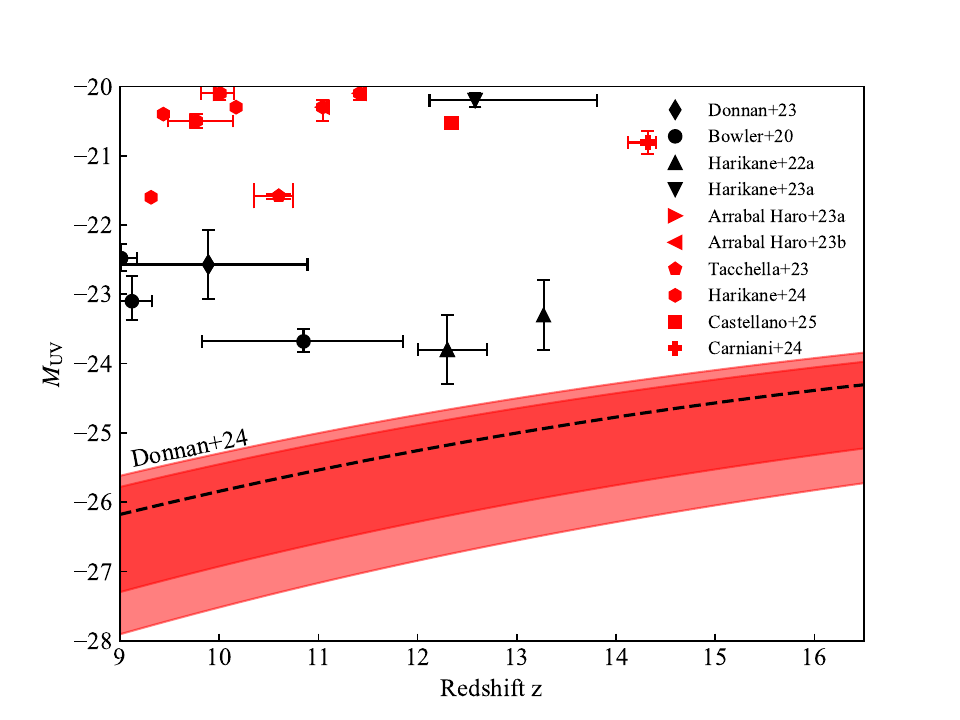}
    \caption{Extreme-value pdf profile (same as \fref{fig:evs}) plotted as a function of redshift, for the double power law luminosity function with parameters given in  \protect\cite{donnan2024jwst}.}\label{fig:evs_dpl_new}
\end{figure}

\mmm

\item \ii{The $f_*$ model} of the luminosity function (as outlined in \cite{harikane2023pure}) was considered as an alternative to the Schechter and DPL models. This model requires the stellar formation rate $f_*$, which we initially assumed to be constant, and a halo mass function (taken to be the Tinker mass function). \fref{fig:evs_f} shows that for constant formation rate, $f_* = 0.02$ produces similar result to that of the  Schechter model, whilst $f_*$ between $0.05$ and $0.15$ produces an extreme-value pdf that is consistent with \textit{JWST} data.

The model predicts that at $z\sim16$, the brightest galaxies will have a UV magnitude around $-23$. Such a prediction depends on the  choice of the halo mass function, which may be rapidly evolving at high redshifts~\cite{donnan2024jwst}. The Tinker mass function is known predict the halo number density accurately at low redshifts ($z \sim 1-2$) and a recalibration to high-$z$ simulations is needed to confirm if the parameter values remain accurate for $z\sim9-16$.

\mmm

\item \ii{Extending the $f_*$ model} to include a dependence on the halo mass was studied in \sref{sec:SFR}. In particular, we considered the models of \cite{tacchella2023jades} (\eref{Tmodel}) and \cite{harikane2022goldrush} (\eref{Hmodel}). The extreme-value predictions are shown in \fref{fig:EVS_ext}, which shows an upward curvature of the bands at high redshifts. The Tacchella model gives a similar result to the constant $f_*=0.15$ model, whilst the Harikane model is comparable to $f_*$ of a few percent, with the predictions diverging at $z\sim16$.  As more high redshift galaxies are observed, the extreme-value formalism will allow us to distinguish whether $f_*$ is constant or halo mass-dependent.

\mmm

\fref{fig:all} summarises the main results of this work: each solid line shows the peak of the extreme-value pdf as a function of redshift, assuming the double power law, constant $f_*$ and variable $f_*$ models that are consistent with \textit{JWST} observation of the brightest high-redshift galaxies.
\begin{figure}
\centering
\includegraphics[width =0.48\textwidth]{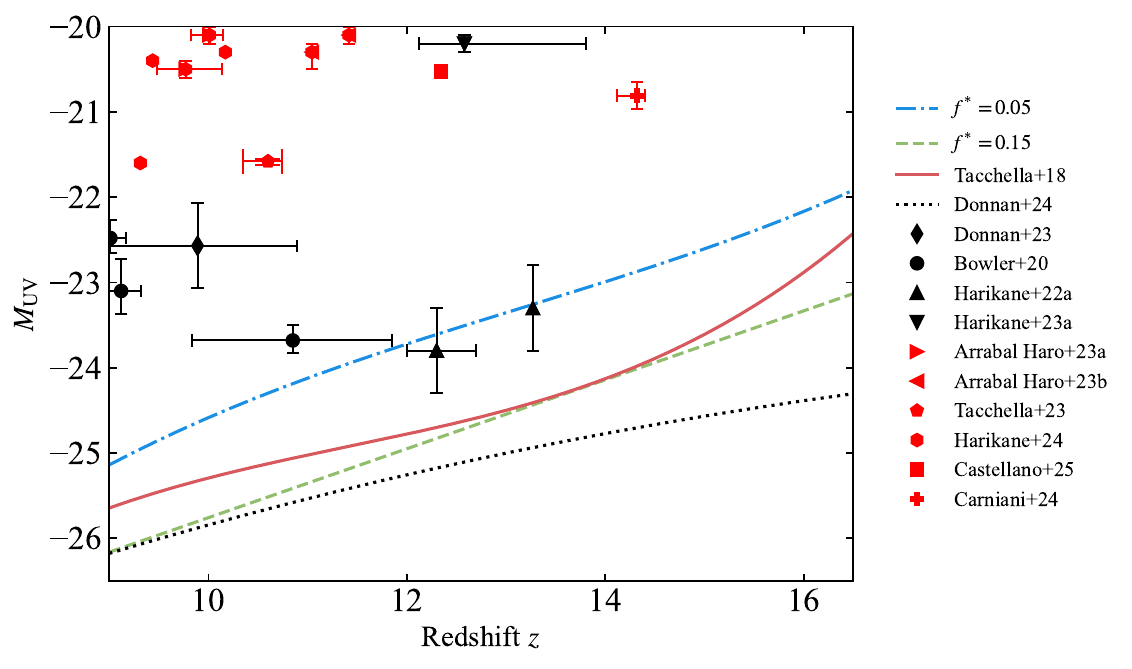}
\caption{A comparison of various EVS models in this work that are consistent with \textit{JWST} data. Each solid lines shows the peak of the Gumbel distribution (percentile bands not shown). The models for the luminosity function $dn/dM\sub{UV}$ are: constant $f_*$ (\eref{eqn:newlum}) with $f_* = 0.05$ (blue) and $f_* = 0.15$ (green), variable $f_*$ (\eref{eqn:dndMfstar}) (pink), and the double-power law model (\eref{eqn:dpl}) assuming the parameters of \protect\cite{donnan2024jwst} (black).}
\label{fig:all}
\end{figure}

\mmm

\item \ii{Uncertainties in the data}. We have taken into account neither the effects of cosmic variance (in the sense of the statistical uncertainty in observation of rare objects at high redshifts) nor Eddington bias 
(\cite{trenti2008cosmic}, \cite{teerikorpi2015eddington}). Both of these could reduce the deduced brightness of extreme galaxies. We leave the calculation of statistical bias to a future work.
\end{itemize}





\section*{Acknowledgements}
For the purpose of open access, the author has applied a Creative Commons Attribution (CC BY) licence to any Author Accepted Manuscript version arising from this submission. Cameron Heather is supported by the Warwick Mathematics Institute Centre for Doctoral Training, and gratefully acknowledges funding from the University of Warwick and the UK Engineering and Physical Sciences Research Council (Grant number: EP/W524645/1). We would like to thank Suraphong Yuma for his useful suggestions.

\section*{Data Availability}

The data used in this article will be shared on reasonable request to the corresponding author.

\bibliographystyle{mnras} 
\bibliography{galaxies}

\appendix
\section{}\label{appendix}

\begin{table*}
    \centering
    \begin{tabular}{||ccccc||}
     \hline
    Luminosity function & Redshift & Peak $M_{\rm UV}$ & $\alpha$ & $ \beta$ \\
     \hline
     DPL & $z\sim9$& $-27.88$ & 27.86 & 0.4960 \\
     (\eref{eqn:dpl})& $z\sim10$& $-28.27$ & 28.21 & 0.5211  \\
     & $z\sim12$& $-29.06$ & 29.09 & 0.6093 \\
     & $z\sim16$& $-31.56$ & 31.58 & 0.9298 \\
     \hline

     DPL & $z\sim9$ & $-26.01$ & 26.02 & 0.3679 \\
      \citep{donnan2024jwst}& $z\sim10$& $-25.69$ & 25.69 & 0.3578  \\
     & $z\sim12$ & $-25.13$ & 25.14 & 0.3392 \\
     & $z\sim16$ & $-24.30$ & 24.31 & 0.3075 \\
     \hline

     Schechter & $z\sim9$& $-23.59$ & 23.59 & 0.09729 \\
       (\eref{eqn:sch_mag})& $z\sim10$& $-23.25$ & 23.24 & 0.09917 \\ 
     & $z\sim12$& $-22.56$ & 22.55 & 0.1030 \\
     & $z\sim16$& $-21.17$ & 21.12 & 0.1112  \\
     \hline

     $f_* = 0.02$ & $z\sim9$& $-23.84$ & 23.82 & 0.1651 \\
       (\eref{eqn:newlum})& $z\sim10$& $-23.43$ & 23.38 & 0.1711 \\
     & $z\sim12$&  $-22.53$ & 22.52 & 0.1751 \\
     & $z\sim16$& $-20.91$ & 20.92 & 0.1907 \\
     \hline

     $f_* = 0.05$ & $z\sim9$& $-24.85$ & 24.81 & 0.1647  \\
     & $z\sim10$& $-24.34$ & 24.37 & 0.1706 \\
     & $z\sim12$& $-23.54$ & 23.52 & 0.1747 \\
     & $z\sim16$& $-21.92$ & 21.92 & 0.1902 \\
     \hline

     $f_* = 0.15$ & $z\sim9$& $-25.96$ & 26.00 & 0.1644 \\
     & $z\sim10$& $-25.56$ & 25.56 & 0.1703 \\
     & $z\sim12$& $-24.75$ & 24.71 & 0.1745 \\
     & $z\sim16$& $-23.13$ & 23.11 & 0.1895 \\
     \hline

     $f_*(M_{\rm h})$ Tacchella & $z\sim9$& $-25.45$ & 25.42 & 0.1277 \\
     (\eref{Tmodel})& $z\sim10$& $-25.15$ & 25.18 & 0.1379 \\
     & $z\sim12$& $-24.65$ & 24.59 & 0.1835 \\
     & $z\sim16$& $-22.42$ & 22.41 & 0.3374 \\
     \hline

     $f_*(M_{\rm h})$ Harikane & $z\sim9$& $-23.84$ & 23.77 & 0.1451\\
     (\eref{Hmodel})& $z\sim10$& $-23.43$ & 23.40 & 0.1981 \\
     & $z\sim12$& $-22.22$ & 22.11 & 0.3296 \\
     & $z\sim16$& $-18.26$ & 18.20 & 0.4495 \\
     \hline

    \end{tabular}
    \caption{The peak magnitude $M_{\rm UV}$, and the  Gumbel parameters $\alpha$ and $\beta$ for the luminosity functions considered in this paper. The notation  $z \sim 9, 10, 12, 16$ refers to the redshift bins $[9,10]$, $[10,11]$, $[12,13]$ and $[16,17]$ respectively.}
    \label{tab:full}
\end{table*}
\tref{tab:full} shows the GEV parameters ($\alpha$ and $\beta$) and peak magnitudes for each luminosity function at various redshifts (see \eref{eqn:GEV}). The parameters have been calculated numerically using least-square fitting, assuming $\gamma=0$ (\ie\ assume that the GEV distribution is the Gumbel type). Alternatively, one can use \eref{eqn:params} to approximate these values.



\bsp	
\label{lastpage}
\end{document}